\documentclass[11pt,a4paper]{scrartcl}

\usepackage{latexsym}
\usepackage{amssymb}
\usepackage{amsmath}
\usepackage[super, sort&compress]{natbib}
\usepackage{alltt}

\usepackage{color}
\definecolor{myred}{rgb}{0.5,0,0}
\definecolor{myblue}{rgb}{0,0,0.75}
\definecolor{mygreen}{rgb}{0,0.5,0}

\usepackage{ifpdf}

\ifpdf
   \usepackage[pdftex]{graphicx}
   \usepackage[pdftex,
               pdftitle={Bayesian estimation of probabilities of default for low default portfolios},
               pdfsubject={Low default portfolio, probability of default, upper confidence bound, Bayesian estimator},
               pdfkeywords={},
               pdfauthor={Dirk Tasche},
               pdfstartview=FitR,
               breaklinks=true,
               colorlinks=true,
               citecolor=myred,
               linkcolor=myblue,
               urlcolor=mygreen]{hyperref}
\else
   \usepackage[dvips]{graphicx}
   \usepackage{hyperref}
   \usepackage{rotating}
\fi

\newtheorem{theorem}{Theorem}
\newtheorem{lemma}{Lemma}
\newtheorem{remark}{Remark}
\newtheorem{example}{Example}
\newtheorem{proposition}{Proposition}

\newtheorem{corollary}{Corollary}
\newtheorem{assumption}{Assumption}

\begin{document}

\renewcommand\bibnumfmt[1]{#1}

\section*{Bayesian estimation of probabilities of default for low default portfolios}

\textbf{Dirk Tasche}\\
{\small{}
is a technical specialist at the Bank of England –- Prudential Regulation Authority (PRA). Before joining
the PRA's predecessor the FSA, he worked for Lloyds Banking Group, Fitch Ratings, and the Deutsche
Bundesbank. Dirk holds a doctorate in probability theory from Berlin University of
Technology. He has published a number of papers on quantitative risk management.\\
The opinions expressed in this paper are those of the author 
and do not necessarily reflect views of the Bank of England.

Bank of England -- PRA, 20 Moorgate, London EC2R 6DA, UK\\
Tel: +44 (0)20 3461 8174; E-mail: dirk.tasche@gmx.net}

\paragraph{Abstract}
The estimation of probabilities of default (PDs) for low default portfolios 
by means of upper confidence bounds is a well established procedure in
many financial institutions. However, there are often discussions within
the institutions or between institutions and supervisors about which confidence level to
use for the estimation. The Bayesian estimator for the PD based on the uninformed, uniform prior distribution is an obvious alternative that avoids the choice of a confidence level. In this paper, we demonstrate that in the case of independent default events the upper confidence bounds can be represented as quantiles of a Bayesian posterior distribution based on a prior that is slightly more conservative than the uninformed prior.  We then describe how to implement the uninformed and conservative Bayesian estimators in the dependent one- and multi-period default data cases and compare their estimates to the upper confidence bound estimates. The comparison leads us to suggest a constrained version of the uninformed (neutral) Bayesian estimator as an alternative to the upper confidence bound estimators.\\[1ex]
\emph{\textbf{Keywords:} Low default portfolio, probability of default, upper confidence bound, Bayesian estimator.}


\section*{Introduction}
\label{tas_sec_0}

The probability of default (PD) per borrower is a core input to modern
credit risk modelling and managing techniques. As such, the appropriateness
of the PD estimations determines the quality of the results of
credit risk models.
Despite the many defaults observed in the recent financial crisis, 
one of the obstacles connected with PD estimates can be the low
number of defaults in the estimation sample because one might experience many years without any default
for good rating grades. Even if some defaults occur in a given year,
the observed default rates might exhibit a high degree of volatility over time. But
even entire portfolios with low or no defaults are not uncommon in practice.
Examples include portfolios with an overall good quality of borrowers
(for example, sovereign or financial institutions portfolios) as well as high exposure but low borrower number
portfolios (for example, specialised lending) and emerging markets portfolios of up to medium
size.

The Basel Committee  might have had
in mind these issues when they wrote in paragraph 451 of the Basel~II 
framework\citep{BaselAccord} ``In general, estimates of PDs, LGDs, 
and EADs are likely to involve unpredictable
errors. In order to avoid over-optimism, a bank must add to its estimates a margin of
conservatism that is related to the likely range of errors. Where methods and data are less
satisfactory and the likely range of errors is larger, the margin of conservatism must be
larger''.

\citet{Pluto&Tasche} suggested an approach to specify the required margin
of conservatism for PD estimates. This method is based on the use of upper confidence
bounds and the so-called most prudent estimation approach. Methods for
building a rating system or a score function on a low default portfolio were proposed by
a number of authors. See \citet{Erlenmaier} for the `rating
predictor' approach and \citet{KennedyMacNameeDelany} and 
\citet{Fernandes&Rocha} for discussions of further alternative approaches.

Although the \citeauthor{Pluto&Tasche} approach to PD estimation
was criticised for delivering too conservative results
\citep{KieferRISK}, it seems to be applied widely
by practitioners nonetheless. Interest in the approach might
have been stimulated to some extent by the UK FSA's
requirement\citep{BIPRU} ``A firm must use a statistical technique to derive the distribution
of defaults implied by the firm's experience, estimating PDs (the
`statistical PD') from the upper bound of a confidence interval
set by the firm in order to produce conservative estimates of PDs $\ldots$''
(4.3.95~R~(2)). The \citeauthor{Pluto&Tasche} approach is also
criticised for the subjectivity it involves as in the multi-period version of
the approach three parameters have to be pre-defined in order to be able 
to come up with a PD estimate. 

However, \citet{Pluto&Tasche2011} suggested
an approach to the estimation of the two correlation parameters that
works reasonably when there is a not too short time-series of default data and
some defaults were recorded in the past. This paper is about
how to get rid of the need to choose a confidence level for the low default
PD estimation.

\citet{Forrest2005} and 
\citet{BenjaminCathcartRyan} 
proposed modifications of the \citeauthor{Pluto&Tasche}
approach in order to facilitate its application and to better control
its inherent conservatism. Other
researchers looked for alternative approaches to statistically based
low default PD estimation. Bayesian methods seem to be most promising.
\citeauthor{KieferJEF}\citep{KieferJEF, KieferJBES, KieferJAE} 
explored in some detail the Bayesian approach 
with prior distributions determined by expert judgment. Clearly,
\citeauthor{KieferRISK}'s approach makes the choice of a confidence level
dispensable. However, this comes at the cost of introducing another source
of subjectivity in the shape of expert judgment. Solutions to this
problem were suggested by \citet{Dwyer.JRMV} and by \citet{Orth} 
who discussed the use of uninformed (uniform) prior distributions and empirical prior distributions
respectively for PD estimation.

In this paper we revisit a comment by \citet{Dwyer.JRMV} on a possible interpretation of
the \citeauthor{Pluto&Tasche} approach in Bayesian terms. We show that indeed
in the independent one-period case the upper confidence bound estimates of PDs are
equivalent to quantiles of the Bayesian posterior distribution of the PDs when
the prior distribution is chosen appropriately conservative (see next section). 
We use the prior distribution
identified this way to define versions of the conservative 
Bayesian estimator of the PD parameter also in the
one-period correlated (third section) and multi-period correlated (fourth section) cases. 

We compare the estimates generated with the conservative Bayesian estimator to
estimates by means of the neutral Bayesian estimator and constrained versions of the 
neutral Bayesian estimator. It turns out that in practice the neutral and the
conservative estimators do not differ very much. In addition, we show that
the neutral estimator can be efficiently calculated in a constrained version (assuming
that the long-run PD is not greater than 10\%) because the constrained estimator produces
results almost identical with the results of the unconstrained estimator. 

The Bayesian approach suggested in this paper is attractive for several reasons:
\begin{itemize}
	\item Its level of conservatism is reasonable.
	\item It makes the often criticised subjective choice of a confidence
level dispensable.
	\item It is sensitive to the presence of correlation in the sense of
	delivering estimates comparable to upper confidence bound estimates
	at levels between 50\% and 75\% for low correlation default time series
	and estimates comparable to 75\% and higher level upper confidence bounds for
	higher correlation default time series.
\end{itemize}

In this paper, we consider only portfolio-wide long-run PD estimates but 
no rating grade-level estimates. 
	How to spread the portfolio-wide estimate on
	sub-portfolios defined by rating grades is discussed by 
	\citet{Pluto&Tasche2011} (`most prudent estimation'),
	by \citet{VanDerBurgt} and by \citet{Tasche2009a} (see Conclusions). 
	The method discussed by \citet{Pluto&Tasche2011}  is
	purely based on sub-portfolio sizes and can lead to hardly different counterintuitive estimates
	for different rating grades. The 
	method proposed by \Citet{VanDerBurgt} and by \citet{Tasche2009a} requires 
	that an estimate of the discriminatory power of the 
	rating system or score function in question is known.

At first glance it might seem questionable to assume that there is one single long-run
PD for an entire portfolio while at the same time trying to estimate long-run
PDs for subportfolios defined by rating grades. However, this assumption can be
justified by taking recourse to the `law of rare events' as presented, e.g.,
in Theorem~6.1 of 
\citet{Durrett}. As a consequence of this theorem,
on a sufficiently large portfolio and as long as the PDs
are not too large, for the distribution of the number of default events 
on the portfolio it does not matter whether the PDs are heterogeneous
or homogeneous.

All calculations for this paper were conducted by means of the statistics software R \citep{RSoftware}.
R-scripts for the calculation of the tables and figures are available upon request from the author.


\section*{One observation period, independent defaults}
\label{sec:OneInd}

Let us recall the low default PD estimation in the independent defaults, one
observation period setting as suggested by \citet{Pluto&Tasche}. The idea is
to use the one-sided upper confidence bound at some confidence level 
$\gamma$ (e.g.\ $\gamma = 50\%$, $\gamma = 75\%$, or $\gamma = 90\%$)
as an estimator of the long-run PD.

\begin{assumption}\label{as:OneInd}
At the beginning of the observation period (in practice often one year) there
are $n > 0$ borrowers in the portfolio. Defaults of borrowers occur independently, and all have
the same probability of default (PD) $0 < \lambda < 1$. At the 
end of the observation period $0 \le k < n$ defaults are observed among the $n$ borrowers.
\end{assumption}

As an example typical for low default portfolios think of Assumption~\ref{as:OneInd} with
$n=1000$ and $k=1$.

What conclusion can we draw from the observation of the number of defaults $k$ 
on the value of the PD $\lambda$? If we have a 
candidate value (an estimate) $\lambda_0$ for $\lambda$ we can statistically test the (Null-)hypothesis
$H_0$ that $\lambda \ge \lambda_0$.

Why $H_0: \lambda \ge \lambda_0$ and not $H^\ast_0: \lambda \le \lambda_0$?
Because if we can reject $H_0$ we have proven (at a usually relatively small type~I error
level, i.e.\
with controlled small probability of erroneous rejection of the Null-hypothesis) that
the alternative $H_1: \lambda < \lambda_0$ is true. Hence we have found an upper bound for the PD
$\lambda$.

It is well-known that under Assumption~\ref{as:OneInd} the number of defaults is binomially distributed
and that the distribution function of the number of defaults can be written in terms of the Beta-distribution 
(see Section 3.2 and Exercise 2.40 of \citet{Casella&Berger}).
\begin{proposition}\label{pr:binomial}
Under Assumption~\ref{as:OneInd} the random number of defaults $X$ in the observation period is binomially
distributed with size parameter $n$ and success probability $\lambda$, i.e.\ we have
\begin{subequations}
\begin{align}
		\mathrm{P}[X \le x] & \ =\ \mathrm{P}_\lambda[X \le x] \ = \ \sum_{\ell=0}^x \bigl(\begin{smallmatrix} n \\ \ell \end{smallmatrix}\bigr) \lambda^\ell\,(1-\lambda)^{n-\ell}, \quad x \in \{0, 1, \ldots, n\}.
\intertext{%
The distribution function of $X$ can be calculated as function of the parameter $\lambda$ as follows:}
\label{eq:beta2}
	\mathrm{P}_\lambda[X \le x] &
	\ = \ 1 - \mathrm{P}[Y \le \lambda] \ = \ \frac{\int_\lambda^1 t^x\, (1-t)^{n-x-1}\, dt} {\int_0^1 t^x\,
(1-t)^{n-x-1}\, dt}, \quad x \in \{0, 1, \ldots, n\},
\end{align}
where $Y$ is Beta-distributed with shape parameters $x+1$ and $n-x$.
\end{subequations}
\end{proposition}
See, e.g., page~623 of \citet{Casella&Berger} for the density and most important properties
of the Beta-distribution.

By means of Proposition~\ref{pr:binomial} we can test $H_0: \lambda \ge \lambda_0$ based on the observed number of defaults $X$ as 
test statistic. If $\mathrm{P}_{\lambda_0}[X \le k] \le \alpha$ for some pre-defined type I error size $0 < \alpha < 1$
($\alpha = 5\%$ is a common choice) we can safely conclude that the outcome of the test is an unlikely event under $H_0$ and 
that, therefore, $H_0$ should be rejected in favour of the alternative $H_1: \lambda < \lambda_0$.
This test procedure is even uniformly most powerful as a consequence of the Karlin-Rubin theorem 
(see Theorem~8.3.17 of \citet{Casella&Berger}) because the binomial distribution has a
monotone likelihood ratio.

If we had $n = 1000$ borrowers in the portfolio at the beginning of the observation period and observed $k=1$ defaults
by the end of the period, testing the Null-hypothesis $H_0: \lambda \ge \lambda_0 = 1\%$ would lead to
\begin{equation}
	\mathrm{P}_{\lambda_0=1\%}[X \le 1]\ =\ 0.05\%.
\end{equation}
Hence under $H_0$ the lower tail probability is clearly less than any commonly accepted type I error size (like 1\% or
5\%) and thus we should reject $H_0$ in favour of the alternative $H_1: \lambda < \lambda_0 = 1\%$.

However, given that the observed default rate was $k/n = 1/1000 = 0.1\%$ a PD estimate of 1\% seems overly 
conservative even if we can be quite sure that the true PD does indeed not exceed 1\% (at least as long as 
we believe that Assumption~\ref{as:OneInd} is justified).

With a view on the fact that the lower tail probability $\mathrm{P}_{\lambda_0=1\%}[X \le 1]$ is much lower than
a reasonable type I error size of -- say -- $\alpha = 5\%$ we might want to refine the arbitrarily chosen upper PD bound 
of $\lambda_0 = 1\%$ by identifying the set of all $\lambda_0$ such that 
\begin{equation}
	\mathrm{P}_{\lambda_0}[X \le 1]\ \le\ \alpha\ =\ 5\%.
\end{equation}
Alternatively we may look for those values of $\lambda_0$ such that 
$H_0: \lambda \ge \lambda_0$ would not have been rejected at $\alpha=5\%$ error level 
for $k$ defaults observed. Technically speaking we then have to find the least $\lambda_0$ such that
still $\mathrm{P}_{\lambda_0}[X \le k] > \alpha$, i.e.\ we want to determine
\begin{subequations}
\begin{equation}\label{eq:inf}
	\lambda_0^\ast \ = \ \inf\{0 < \lambda_0 < 1: \mathrm{P}_{\lambda_0}[X \le k] > \alpha\}.
\end{equation}
Under Assumption \ref{as:OneInd}, by continuity, $\lambda_0^\ast$ solves the equation
\begin{equation}\label{eq:theta2}
	\sum_{\ell=0}^k \bigl(\begin{smallmatrix} n \\ \ell \end{smallmatrix}\bigr) (\lambda_0^\ast)^\ell\,
	(1-\lambda_0^\ast)^{n-\ell} \ = \ \mathrm{P}_{\lambda_0^\ast}[X \le k] \ = \ \alpha.
\end{equation}
Equation \eqref{eq:beta2} implies that the solution of \eqref{eq:theta2} is the $(1-\alpha)$-quantile of a related
Beta-distribution:
\begin{equation}\label{eq:explicit}
	\lambda_0^\ast \ =\ q_{1-\alpha}(Y) \ =\ \min\{y: \mathrm{P}[Y\le y]\ge 1-\alpha\},
\end{equation}
\end{subequations}
where $Y$ is Beta-distributed with shape parameters $k+1$ and $n-k$. If we again consider the
case $n=1000$, $k=1$ and $\alpha = 5\%$ we obtain from \eqref{eq:explicit} that
\begin{subequations}
\begin{equation}
	\lambda_0^\ast \ = \ 0.47\%.
\end{equation}
This estimate of the PD $\lambda$ is much closer to the observed default rate of $0.1\%$ but still -- from a practitioner's
point of view -- very conservative. Let us see how the estimate changes when we choose much higher type I error sizes of
25\% and 50\% respectively (note that such high type~I error levels would not be acceptable from a test-theoretic perspective).
With $n=1000$, $k=1$ and $\alpha = 25\%$ we obtain
\begin{equation}
	\lambda_0^\ast \ = \ 0.27\%.
\end{equation}
The choice $n=1000$, $k=1$ and $\alpha = 50\%$ gives
\begin{equation}
	\lambda_0^\ast \ = \ 0.17\%.
\end{equation}
\end{subequations}
These last two estimates appear much more appropriate for the purpose of credit
pricing or impairment forecasting although we 
have to acknowledge that due to the independence condition 
of Assumption \ref{as:OneInd} we are clearly ignoring cross-sectional and over time correlation effects (which will be discussed in the following two sections).

\begin{remark}
While the independence assumption appears unrealistic in the context of long-run PD estimation, it 
might be appropriate for the estimation of loss given default (LGD) or conversion factors 
for exposure at default (EAD). This comment applies to the situation where only zero 
LGDs or conversion factors were historically observed. The low default estimation method 
could then be used for estimating the probability of
a positive realisation of an LGD or conversion factor. Combined with the conservative
assumption that a positive realisation would be 100\%, such a probability of
a positive realisation would give a conservative LGD or conversion factor estimate.
\end{remark}

Before we discuss what type I error levels are appropriate for the estimation of long-run PDs by way of
\eqref{eq:inf} and solve this issue by taking recourse to Bayesian estimation methods, let us summarize 
what we have achieved so far.

We have seen that -- under Assumption \ref{as:OneInd} -- reasonable upper bounds for the long-run PD $\lambda$
can be determined by identifying the set of estimates $\lambda_0$ such that the hypotheses $H_0: \lambda \ge \lambda_0$
are rejected at some pre-defined type I error level $\alpha$. By \eqref{eq:inf} and \eqref{eq:explicit} this
set has the shape of an half-infinite interval $[\lambda_0^\ast, \infty)$. Equivalently, one could say
that there is an half-infinite interval $(-\infty, \lambda_0^\ast]$ of all the values of 
$\lambda_0$ such that the hypotheses $H_0: \lambda \ge \lambda_0$
are accepted at the type I error level $\alpha$. By the general duality theorem
for statistical tests and  confidence sets (see Theorem~9.2.2 of \citet{Casella&Berger})
we have `inverted' the family of type I error level $\alpha$ tests specified by \eqref{eq:inf} to arrive at 
a one-sided confidence interval  $(-\infty, \lambda_0^\ast]$ at level $\gamma = 1-\alpha$ for the PD $\lambda$
which is characterised by the upper confidence bound $\lambda_0^\ast$. This observation does
not depend on any distributional assumption like Assumption \ref{as:OneInd}.

\begin{proposition}\label{pr:upper}
For any fixed confidence level $0 < \gamma < 1$, the number $\lambda_0^\ast(\gamma)$ defined 
by \eqref{eq:inf} with $\alpha = 1 - \gamma$ represents an upper confidence bound 
at level $\gamma$ for the 
PD~$\lambda$.
\end{proposition}
By Theorem 9.3.5 of \citet{Casella&Berger}, the confidence interval $(-\infty, \lambda_0^\ast]$ is the
\emph{uniformly most accurate} confidence interval among all one-sided confidence intervals at level $\gamma$
for $\lambda$.
Together with \eqref{eq:explicit} Proposition~\ref{pr:upper} implies the following convenient representation
of the upper confidence bounds.
\begin{corollary}\label{co:upper}
Under Assumption \ref{as:OneInd}, for any fixed confidence level $0 < \gamma < 1$, an upper confidence bound
$\lambda_0^\ast(\gamma)$ for the PD $\lambda$ at level $\gamma$ can be calculated by 
\eqref{eq:explicit} with $\alpha = 1 - \gamma$.
\end{corollary}
By Corollary \ref{co:upper}, the upper confidence bounds for $\lambda$ are just the $\gamma$-quantiles
of a Beta-distribution with shape parameters $k+1$ and $n-k$. This observation makes it possible to identify
the upper confidence bounds with Bayesian upper credible bounds for a specific non-uniform prior distribution
of $\lambda$. See Section~9.2.4 of \citet{Casella&Berger} for a discussion of the conceptual differences
between classical confidence sets and Bayesian credible sets. The following result 
is a generalization of \citet{Dwyer.JRMV} (Appendix C).

\begin{theorem}[Bayesian posterior distribution of PD]\label{th:Bayesian}
Under Assumption \ref{as:OneInd}, assume in addition that the PD $0< \lambda <1$ is the realisation of
a random variable $\Lambda$ with unconditional (prior) distribution
\begin{subequations}
\begin{equation}\label{eq:apriori}
	\pi\bigl((0,\lambda]\bigr) = \int_0^\lambda \frac{d u}{1-u} = -\log(1-\lambda), \ 0< \lambda <1.
\end{equation}
Denote by $X$ the number of defaults observed at the end of the observation period. Then the conditional
(posterior) distribution of the PD $\Lambda$ given $X$ is
	\begin{equation}\label{eq:representation}
	\mathrm{P}[\Lambda \le \lambda\,|\,X = k] \ =\ 
	\frac{\int\limits_0^\lambda \ell^k\,(1-\ell)^{n-k-1} d\,\ell}{\int\limits_0^1 \ell^k\,(1-\ell)^{n-k-1} d\,\ell},
	\quad k \in\{0, 1, \ldots, n-1\},
\end{equation}
i.e.\ conditional on $X=k$ the distribution of $\Lambda$ is a Beta-distribution with shape parameters 
$k+1$ and $n-k$.
\end{subequations}
\end{theorem}
Note that $\pi$ is not a probability distribution as $\pi\bigl((0,1)\bigr) = \infty$. However,
in a Bayesian context working with \emph{improper} prior distributions is common as the prior 
distribution is only needed to reflect differences in the initial subjective presumptions on
the likelihoods of the parameters to be estimated. Due to the condition $k < n$ from Assumption \ref{as:OneInd},
the posterior distribution of $\Lambda$ turns out to be a proper probability distribution.

\paragraph{Proof of Theorem~\ref{th:Bayesian}.} By Proposition \ref{pr:binomial}, since $k < n$ Equation
\eqref{eq:representation} is the result of the following calculation:
\begin{align*}
	\mathrm{P}[\Lambda \le \lambda\,|\,X = k] & = 
	\frac{\mathrm{P}[\Lambda \le \lambda, X = k]}{\mathrm{P}[X = k]}\notag\\[1ex]
	& = \frac{\int\limits_0^\lambda \mathrm{P}[X=k\,|\,\Lambda=\ell]\,\frac{\partial \pi((0, \ell])}
	{\partial \ell}\, d \ell}
	{\int\limits_0^1 \mathrm{P}[X=k\,|\,\Lambda=\ell]\,\frac{\partial \pi((0, \ell])}
	{\partial \ell}\, d \ell}\notag\\[1ex]
	& =\frac{\int\limits_0^\lambda \bigl(\begin{smallmatrix} n \\ k \end{smallmatrix}\bigr) \ell^k\,(1-\ell)^{n-k}\,\frac{d\,\ell}{1-\ell}}
	{\int\limits_0^1 \bigl(\begin{smallmatrix} n \\ k \end{smallmatrix}\bigr) 
	\ell^k\,(1-\ell)^{n-k}\,\frac{d\,\ell}{1-\ell}} \notag\\[1ex]
	&= \frac{\int\limits_0^\lambda \ell^k\,(1-\ell)^{n-k-1} d\,\ell}{\int\limits_0^1 \ell^k\,(1-\ell)^{n-k-1} d\,\ell}.
\end{align*}
This proves the assertion. \hfill \textbf{q.e.d.}

At first glance, the prior distribution \eqref{eq:apriori} with the singularity in $\lambda =1$ seems 
heavily biased towards the higher potential values of $\lambda$. Due to this conservative bias, it makes sense
to call the distribution \eqref{eq:apriori} a \emph{conservative} prior distribution. In any case,
it is interesting to note that the density $\lambda \mapsto \frac 1{1-\lambda}$ of the prior distribution 
\eqref{eq:apriori} is increasing. This is a feature the conservative prior has in common
with the characteristic densities of spectral risk
measures, a special class of coherent risk measures \citep{Acerbi_Spectral, Tasche2002}. We will see
below that the conservative shift induced by the prior distribution \eqref{eq:apriori} is actually
quite moderate.

By definition, in a Bayesian setting a credible upper bound of a parameter 
is a quantile of the posterior distribution of the parameter. By Corollary \ref{co:upper} and
Theorem \ref{th:Bayesian}, since both the classical confidence bounds and the Bayesian credible bounds 
are quantiles of the same Beta-distribution, hence we can state the 
following result:

\begin{corollary}\label{co:keyresult}
Under Assumption \ref{as:OneInd}, if the Bayesian prior distribution of the PD $\lambda$ is given by \eqref{eq:apriori}
then the classical one-sided upper confidence bounds at level $0 < \gamma < 1$ and the Bayesian one-sided
upper credible bound of $\lambda$ coincide and are determined by \eqref{eq:explicit} with $\alpha = 1-\gamma$.
\end{corollary}

Corollary \ref{co:keyresult} is a key result of this paper. We already knew from \eqref{eq:explicit} that the
upper confidence bounds suggested by \citet{Pluto&Tasche} as conservative estimates of the PD can be determined
as quantiles of a Beta-distribution. However, Corollary \ref{co:keyresult} identifies this specific Beta-distribution
as a Bayesian posterior distribution of the PD for the conservative prior distribution \eqref{eq:apriori}.

In order to assess the extent of conservatism induced by the prior distribution~\eqref{eq:apriori} we
introduce a family of uniform prior distributions as described in the following proposition.

\begin{proposition}\label{pr:uniform}
Under Assumption \ref{as:OneInd}, let the Bayesian prior distribution of the PD $\lambda$ be given by the
uniform distribution on the interval $(0, u)$ for some $0 < u \le 1$. 
Denote by $X$ the number of defaults observed at the end of the observation period. Then the conditional
(posterior) distribution of the PD given $X$ is specified by the density $f$ with
\begin{equation}\label{eq:Bayes.restricted}
	f(\lambda) \ = \ \begin{cases}
	0, & 1 > \lambda \ge u,\\
	\frac{b_{k+1, n-k+1}(\lambda)}{\mathrm{P}[Y \le u]}, & u > \lambda > 0,
\end{cases}
\end{equation}
where $b_{k+1, n-k+1}$ denotes the density of the Beta-distribution with shape parameters $k+1$ and $n-k+1$ 
and $Y$ is a random variable with this distribution.
\end{proposition}
\textbf{Proof.} The calculation for this proof is rather similar to the calculation in the proof of
Theorem~\ref{th:Bayesian}. Denote by $\Lambda$ a random variable with uniform distribution on $(0, u)$
which in the Bayesian context is associated with the PD. Then we have for $0 < \lambda < 1$
\begin{align}
	\mathrm{P}[\Lambda \le \lambda\,|\,X = k] & = 
	\frac{\mathrm{P}[\Lambda \le \lambda, X = k]}{\mathrm{P}[X = k]}\notag\\[1ex]
	& = \frac{\int_0^{\min(u,\lambda)} \mathrm{P}[X=k\,|\,\Lambda=\ell]\,d\, \ell}
	{\int_0^{u} \mathrm{P}[X=k\,|\,\Lambda=\ell]\, d\, \ell}\notag\\[1ex]
	& =\frac{\int_0^{\min(u,\lambda)} \bigl(\begin{smallmatrix} n \\ k \end{smallmatrix}\bigr) \ell^k\,(1-\ell)^{n-k}\,d\,\ell}
	{\int_0^{u} \bigl(\begin{smallmatrix} n \\ k \end{smallmatrix}\bigr) 
	\ell^k\,(1-\ell)^{n-k}\,d\,\ell} \notag\\[1ex]
	&= \frac{\int_0^{\min(u,\lambda)} b_{k+1, n-k+1}(\lambda)\, d\,\ell}
	{\int_0^{u} b_{k+1, n-k+1}(\lambda)\, d\,\ell}.
	\label{eq:beta.restricted}
\end{align}
Equation~\eqref{eq:beta.restricted} implies \eqref{eq:Bayes.restricted}. \hfill \textbf{q.e.d.}

Observe that in the special case $u=1$ of Proposition~\ref{pr:uniform} the posterior distribution
of the PD is the Beta-distribution with shape parameters $k+1$ and $n-k+1$ as is well-known
from textbooks like \citet{Casella&Berger} (see Example 7.2.14).

The most natural estimator associated with a Bayesian posterior distribution is its mean. We determine the mean
associated with the conservative prior \eqref{eq:apriori} in the following proposition. 
It is also of interest to consider the Bayesian estimators associated with the uniform distributions
introduced in Proposition~\ref{pr:uniform}. In particular, the uniform distribution on $(0,1)$  
is the natural uninformed (or neutral) prior for probability parameters.

\begin{proposition}\label{pr:Bayes.est}
Under Assumption \ref{as:OneInd}, if the Bayesian prior distribution of the PD $\lambda$ is given by \eqref{eq:apriori}
then the mean $\lambda_1^\ast$ of the posterior distribution is given by
\begin{subequations}
\begin{equation}\label{eq:conservative}
	\lambda_1^\ast \ =\ \frac{k+1}{n+1}.
\end{equation}
$\lambda_1^\ast$ is called the \emph{conservative Bayesian estimator} of the PD $\lambda$. 
If the Bayesian prior distribution of the PD $\lambda$ is given by the uniform distribution on $(0,u)$
for some $0 < u \le 1$
then the mean $\lambda_2^\ast(u)$ of the posterior distribution is given by
\begin{equation}\label{eq:neutral}
	\lambda_2^\ast(u) \ =\ 
	\frac{(k+1)\,\mathrm{P}[Y_{k+2, n-k+1}\le u]}{(n+2)\,\mathrm{P}[Y_{k+1, n-k+1}\le u]},
\end{equation}
\end{subequations}
where $Y_{\alpha, \beta}$ denotes a random variable which is Beta-distributed
with paramaters $\alpha$ and $\beta$.
$\lambda_2^\ast(u)$ is called the \emph{$(0,u)$-constrained neutral Bayesian estimator} of the PD $\lambda$.
For $u = 1$, we obtain the (unconstrained) \emph{neutral Bayesian estimator} $\lambda_2^\ast(1)$.
\end{proposition}
\textbf{Proof.} According to Theorem \ref{th:Bayesian}, the posterior distribution of the PD associated
with the conservative prior distribution is the Beta-distribution with parameters $k+1$ and $n-k$. As the mean
of this Beta-distribution is $\frac{k+1}{n+1}$ this proves \eqref{eq:conservative}. For \eqref{eq:neutral} 
we can compute
\begin{align}
	\mathrm{E}[\Lambda\,|\,X = k] & = 
	\frac{\int\limits_0^u \ell\,\mathrm{P}[X=k\,|\,\Lambda=\ell]\,d\, \ell}
	{\int\limits_0^{u} \mathrm{P}[X=k\,|\,\Lambda=\ell]\, d\, \ell}\notag\\[1ex]
	& =\frac{\int\limits_0^u \bigl(\begin{smallmatrix} n \\ k \end{smallmatrix}\bigr) \ell^{k+1}\,(1-\ell)^{n-k}\,d\,\ell}
	{\int\limits_0^{u} \bigl(\begin{smallmatrix} n \\ k \end{smallmatrix}\bigr) 
	\ell^k\,(1-\ell)^{n-k}\,d\,\ell} \notag\\[1ex]
	&=  \frac{(k+1)\,\int\limits_0^u b_{k+2, n-k+1}(\ell)\, d\,\ell}
	{(n+2)\,\int\limits_0^{u} b_{k+1, n-k+1}(\ell)\, d\,\ell}.
	\notag
\end{align}
This completes the proof.  \hfill \textbf{q.e.d.} 

Observe that in the special case $u=1$ of Proposition~\ref{pr:Bayes.est} the neutral Bayesian estimator
is given by 
\begin{equation}\label{eq:neutral.1}
	\lambda_2^\ast(1) \ =\ \frac{k+1}{n+2}.
\end{equation}
The constrained neutral Bayesian estimator $\lambda_2^\ast(u)$ is differentiable with respect to $u$ in the
open interval $(0,1)$. This follows from the following easy-to-prove lemma:
\begin{lemma}\label{le:diff}
Let $h(\lambda)$, $h: (0,1) \to (0,\infty)$ be a continuous function. Then the function
\begin{subequations}
\begin{align}
	H(u) & \ = \ \frac{\int_0^u \lambda\,h(\lambda)\,d\lambda}%
									{\int_0^u h(\lambda)\,d\lambda}
\intertext{is continuously differentiable with}
	H'(u) & \ = \ h(u)\,\frac{\int_0^u (u-\lambda)\,h(\lambda)\,d\lambda}%
								{\bigl(\int_0^u h(\lambda)\,d\lambda\bigr)^2} \ >\ 0.
\end{align}
\end{subequations}
\end{lemma}
With $h(\lambda) = u^k\,(1-u)^{n-k}$ Lemma~\ref{le:diff} immediately implies that $\lambda_2^\ast(u)$ is increasing in $u$ as one would
intuitively expect.
When comparing $\frac k n$, the naive estimator of the PD under Assumption \ref{as:OneInd}, to the Bayesian
estimators $\lambda_1^\ast$ and $\lambda_2^\ast(u)$, we can therefore notice the following inequalities:
\begin{equation}\label{eq:most}
\left.\begin{gathered}
 \frac k n \ < \ \frac{k+1}{n+1}\ =\ \lambda_1^\ast,\\
 \lambda_2^\ast(u) \ \le \ \frac{k+1}{n+2}\ =\ \lambda_2^\ast(1)
 \ < \ \frac{k+1}{n+1}\ =\ \lambda_1^\ast,\\
 \frac k n \ \le \ \lambda_2^\ast(1) \ = \ \frac{k+1}{n+2}\quad \iff \quad 2\,k\ \le\ n. 
\end{gathered}\ \right\} 	
\end{equation}
\begin{table}[t!p]
\caption{Different PD estimates under
Assumption \ref{as:OneInd} with $k=1$. Upper confidence bounds according to Corollary~\ref{co:keyresult}.
Naive estimator is $\frac k n$. Conservative and neutral Bayesian estimators 
according to Proposition~\ref{pr:Bayes.est}.}
\label{tab:1}
\begin{center}
\begin{tabular}{|l|c|c|c|c|c|}
\hline
Estimator & $n=125$ & 250 & 500 & 1000 & 2000 \\ \hline \hline
Naive & 0.8\% & 0.4\% & 0.2\% & 0.1\% & 0.05\% \\ \hline
50\% upper confidence bound & 1.339\% & 0.6704\% & 0.3354\% & 0.1678\% & 0.0839\% \\ \hline
75\% upper confidence bound & 2.1396\% & 1.0734\% & 0.5376\% & 0.269\% & 0.1346\% \\ \hline
90\% upper confidence bound & 3.076\% & 1.5469\% & 0.7757\% & 0.3884\% & 0.1943\% \\ \hline
Neutral Bayesian on (0, 0.025) & 1.1785\% & 0.7655\% & 0.3983\% & 0.1996\% & 0.0999\% \\ \hline
Neutral Bayesian on (0, 0.05) & 1.5233\% & 0.7935\% & 0.3984\% & 0.1996\% & 0.0999\% \\ \hline
Neutral Bayesian on (0, 0.1) & 1.5746\% & 0.7937\% & 0.3984\% & 0.1996\% & 0.0999\% \\ \hline
Neutral Bayesian on (0,1) & 1.5748\% & 0.7937\% & 0.3984\% & 0.1996\% & 0.0999\% \\ \hline
Conservative Bayesian & 1.5873\% & 0.7968\% & 0.3992\% & 0.1998\% & 0.1\% \\ \hline
\end{tabular}
\end{center}
\end{table}

Hence, the conservative Bayesian estimator is indeed more conservative than the naive estimator
and the neutral Bayesian estimators. We conclude this section with a numerical example 
(Table~\ref{tab:1}), comparing 
the three estimators from \eqref{eq:most} and 
the three upper confidence bounds at 50\%, 75\%, and 90\% levels. From this example, 
some conclusions can be drawn: 
\begin{itemize}
\item Under the assumption of independent defaults, the Bayesian estimators 
tend to assume values between the 50\% and 75\% upper confidence bounds.
Hence, choosing confidence levels between 50\% and 75\% seems plausible. This conclusion
will be confirmed later in the penultimate section.
\item However, as we will see in the following two sections, 
example calculations for the dependent case indicate that then the Bayesian estimators tend
to assume values between 75\% and 90\% upper confidence bounds.
\item The difference between the neutral and the conservative Bayesian estimators
is relatively small and shrinks even more for larger $n$. This observation holds
in general as will be demonstrated in the next sections.
\end{itemize}


\section*{One observation period, correlated defaults}
\label{sec:OneDep}

In this section, we replace the unrealistic assumption of defaults occuring independently
by the assumption that default correlation is caused by one factor dependence as in the Basel II
credit risk model\citep{BaselExplanatoryNote}.

\paragraph{Notation.} $\Phi$ denotes the standard normal distribution function.
$\Phi_2$ denotes the bivariate normal distribution with standardised marginals.
$\varphi$ denotes the standard normal density function $\varphi(s) = \frac{e^{- s^2/2}}{\sqrt{2\,\pi}}$.

\begin{assumption}\label{as:OneCorr}
At the beginning of the observation period there
are $n > 1$ borrowers in the portfolio. All defaults of borrowers have
the same probability of default (PD) $0 < \lambda < 1$. 
The event $D_i$ `borrower $i$ defaults during the observation period' can be
described as follows: 
\begin{equation}\label{eq:DefaultEvent}
	D_i \ = \ \{\sqrt{\varrho}\,S + \sqrt{1-\varrho}\,\xi_i \le \Phi^{-1}(\lambda)\},
\end{equation}
where $S$ and $\xi_i$, $i = 1, \ldots, n$, are independent and standard normal. $S$ is called
\emph{systemic factor}, $\xi_i$ is the \emph{idiosyncratic factor} relating to borrower $i$. The
parameter $0 \le \varrho < 1$ is called \emph{asset correlation}.
At the 
end of the observation period $0 \le k < n$ defaults are observed among the $n$ borrowers.
\end{assumption}

By \eqref{eq:DefaultEvent}, in the case $\varrho > 0$ the default events are no longer independent:
\begin{equation}\label{eq:dep_events}
	\mathrm{P}[\text{Borrowers \emph{i} and \emph{j} default}] = \mathrm{P}[D_i \cap D_j] = \Phi_2\bigl(\Phi^{-1}(\lambda),\Phi^{-1}(\lambda); \varrho\bigr)	> \lambda^2 =
	\mathrm{P}[D_i]\,\mathrm{P}[D_j].
\end{equation} 
We exclude the case $\varrho = 1$ from Assumption~\ref{as:OneCorr} because it corresponds
to the situation where there is only one borrower.

Without independence, Proposition \ref{pr:binomial} does no longer apply. However, the following
easy-to-prove modification holds.

\begin{proposition}\label{pr:corrbinomial}
Under Assumption~\ref{as:OneCorr} the random number of defaults $X$ in the observation period is 
\emph{correlated binomially distributed} with size parameter $n$, success probability $\lambda$, and 
asset correlation parameter $0 \le \varrho < 1$. The distribution of $X$ can be represented as
follows:
\begin{subequations}
\begin{align}\label{eq:corrbinomial}
	\mathrm{P}[X \le k] & = \int\limits_{-\infty}^\infty \varphi(y)\,\sum_{i=0}^k
\left(\begin{smallmatrix}
  n\\ i
\end{smallmatrix}\right)
\,G(\lambda,\varrho, y)^i\,(1-G(\lambda,\varrho, y))^{n-i}\,d\,y, \\
G(\lambda,\varrho, y) & =
   \Phi\Big(\frac{\Phi^{-1}(\lambda)-\sqrt{\varrho}\,y}{\sqrt{1-\varrho}}\Big) = \mathrm{P}[D\,|\,S=y].
   \label{eq:G}
\end{align}
The mean and the variance of $X$ are given by 
\begin{equation}\label{eq:char}
	\begin{split}
	\mathrm{E}[X] & = n\,\lambda, \\ 
\mathrm{var}[X] & = n\,(\lambda - \lambda^2) + 
n\,(n-1)\,\bigl(\Phi_2\bigl(\Phi^{-1}(\lambda),\Phi^{-1}(\lambda); \varrho\bigr)	- \lambda^2\bigr).
	\end{split}
\end{equation}
\end{subequations}
\end{proposition}
$\mathrm{P}[X \le k]$ can be efficiently calculated by numerical integration. Alternatively, one can make use
of a representation of $\mathrm{P}[X = k]$ by the distribution function of an $n$-variate normal distribution:
\begin{equation}
	\mathrm{P}[X = k] \ = \ \left(\begin{smallmatrix}
  n\\ k
\end{smallmatrix}\right) \, \mathrm{P}\bigl[Z_1 \le \Phi^{-1}(\lambda), \ldots, Z_k \le \Phi^{-1}(\lambda),
											Z_{k+1} > \Phi^{-1}(\lambda), Z_n > \Phi^{-1}(\lambda)\bigr],
\end{equation}
where $(Z_1, \ldots, Z_n)$ is multi-variate normal with $Z_i \sim \mathcal{N}(0,1)$, $i = 1, \ldots, n$, and
$\mathrm{corr}[Z_i, Z_j] = \varrho$, $i \not= j$. 

\begin{figure}[t!p]
\caption{Binomial and correlated binomial distributions with same size
and sucess probability parameters.}
\label{fig:1}
\begin{center}
\ifpdf
	\includegraphics[width=15cm]{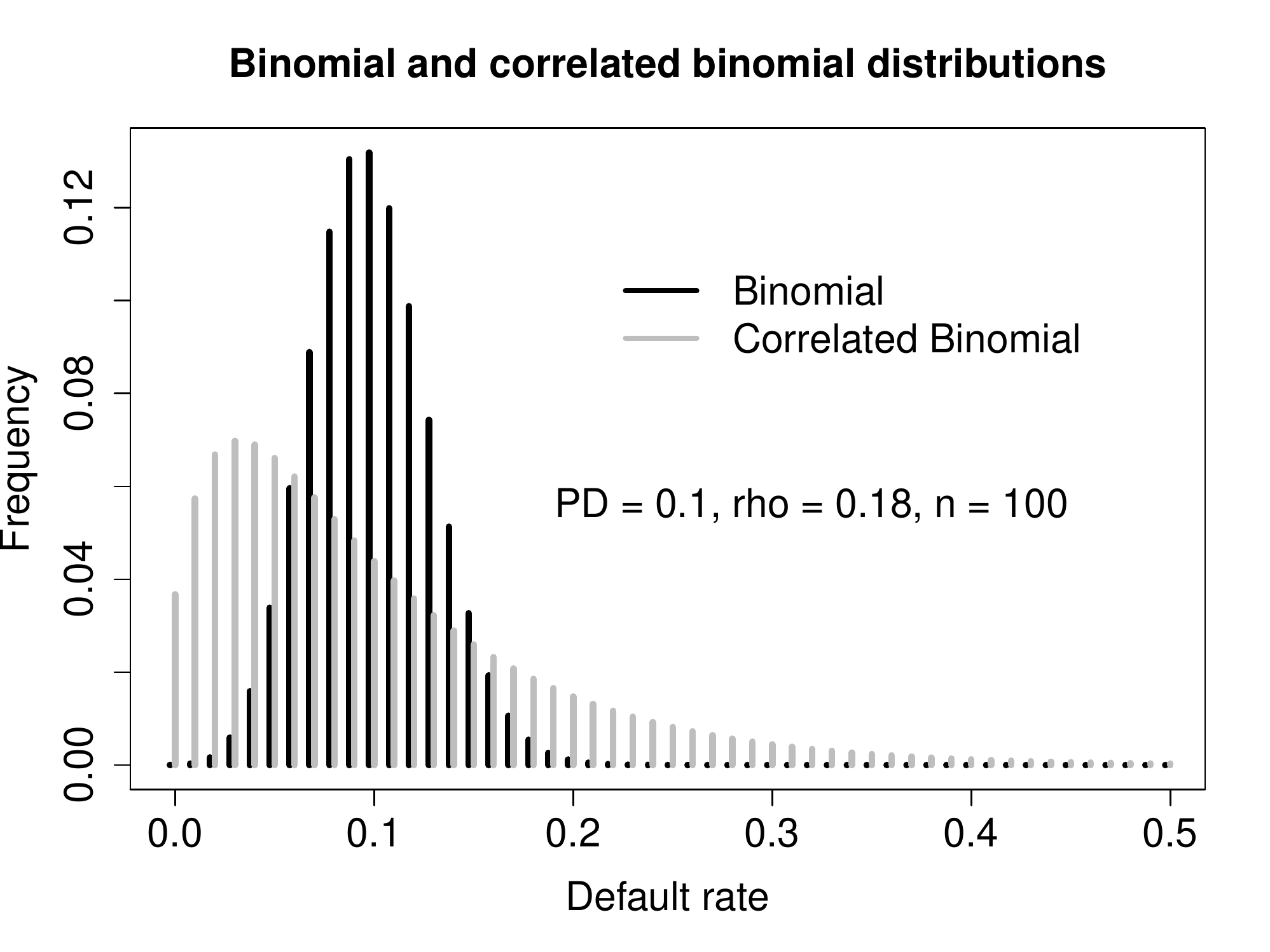}
\else
\begin{turn}{270}
	\includegraphics[width=15.0cm]{Binomial_vs_Vasicek.ps}
\end{turn}
\fi
\end{center}
\end{figure}

Figure~\ref{fig:1} demonstrates the impact of introducing
correlation as by Assumption~\ref{as:OneCorr} on the binomial distribution. The variance of the distribution
is much enlarged (as can be seen from \eqref{eq:char} and \eqref{eq:dep_events}), 
and so is the likelihood of assuming large or small values at some distance from the mean.

With regard to estimators for the PD $\lambda$ from Assumption~\ref{as:OneCorr}, Equation~\eqref{eq:inf}
represents the general approach to upper confidence bound estimators, i.e.\ 
Proposition~\ref{pr:upper} still holds in the more general 'correlated' 
context of Assumption~\ref{as:OneCorr}. Under Assumption~\ref{as:OneCorr},
however, Corollary~\ref{co:upper} no longer applies. Neither does apply Proposition~\ref{pr:Bayes.est}
such that there is no easy way of calculating the Bayesian estimates in the case
of correlated defaults. Instead the upper confidence bound estimates and the Bayesian
estimates have to be calculated by numerical procedures involving one- and 
and two-dimensional numerical integration and numerical root finding (for the confidence bounds) -- as
noted in the following proposition.

\begin{proposition}\label{pr:correlated}
Let $\mathrm{P}_\lambda[X=k] =	\int\limits_{-\infty}^\infty \varphi(y)\,
\left(\begin{smallmatrix} n\\ k \end{smallmatrix}\right)\,
G(\lambda,\varrho, y)^k\,(1-G(\lambda,\varrho, y))^{n-k}\,d\,y$ with 
the function $G(\cdot)$ being defined as in Proposition~\ref{pr:corrbinomial}.
Under Assumption \ref{as:OneCorr}, then we have the following estimators for
the PD parameter $\lambda$:
\begin{itemize}
	\item[(i)] 
For any fixed confidence level $0 < \gamma < 1$, the upper confidence bound
$\lambda_0^\ast(\gamma)$ for the PD $\lambda$ at level $\gamma$ can be calculated by 
equating the right-hand side of \eqref{eq:corrbinomial} to $1-\gamma$ and solving
the resulting equation for $\lambda$.
\item[(ii)] If the Bayesian prior distribution of the PD $\lambda$ is defined by \eqref{eq:apriori}
then the mean $\lambda_1^\ast$ of the posterior distribution is given by
\begin{subequations}
\begin{equation}\label{eq:conservative.corr}
	\lambda_1^\ast \ =\ \frac{\int_0^1 \frac{\lambda\,\mathrm{P}_\lambda[X=k]}{1-\lambda}\, d\lambda}%
			{\int_0^1 \frac{\mathrm{P}_\lambda[X=k]}{1-\lambda}\, d\lambda}.	
\end{equation}
In particular, the integrals in the numerator and the denominator of the
right-hand side of \eqref{eq:conservative.corr} are finite.
$\lambda_1^\ast$ is called the \emph{conservative Bayesian estimator} of the PD $\lambda$. 
\item[(iii)] If the Bayesian prior distribution of the PD $\lambda$ is 
uniform on $(0,u)$ for some $0 < u \le 1$ then the mean $\lambda_2^\ast(u)$ 
of the posterior distribution is given by
\begin{equation}\label{eq:neutral.corr}
	\lambda_2^\ast(u) \ =\ \frac{\int_0^u \lambda\,\mathrm{P}_\lambda[X=k]\, d\lambda}%
			{\int_0^u \mathrm{P}_\lambda[X=k]\, d\lambda}.
\end{equation}
$\lambda_2^\ast(u)$ is called the \emph{$(0,u)$-constrained neutral Bayesian estimator} of the PD $\lambda$. 
For $u = 1$, we obtain the (unconstrained) \emph{neutral Bayesian estimator} $\lambda_2^\ast(1)$.
\end{subequations}
\end{itemize}
\end{proposition}
\textbf{Proof.} Only the statement that the integrals in \eqref{eq:conservative.corr}
are finite is not obvious. Observe that both for $a=0$ and $a=1$ we have
\begin{equation*}
	\int_0^1 \frac{\lambda^a\,\mathrm{P}_\lambda[X=k]}{1-\lambda}\, d\lambda \ \le\
	\int_0^1 \frac{\lambda^a\,
	\bigl(1-\mathrm{E}\bigl[\Phi\bigl(\frac{\Phi^{-1}(\lambda)-\sqrt{\varrho}\,Y}%
	{\sqrt{1-\varrho}}\bigr)\bigr]\bigr)}{1-\lambda}\, d\lambda,
\end{equation*}
where $Y$ denotes a standard normal random variable. It is, however, a well-known fact
that $\mathrm{E}\bigl[\Phi\bigl(\frac{\Phi^{-1}(\lambda)-\sqrt{\varrho}\,Y}%
	{\sqrt{1-\varrho}}\bigr)\bigr] = \lambda$. This implies
$\int_0^1 \frac{\lambda^a\,\mathrm{P}_\lambda[X=k]}{1-\lambda}\, d\lambda < \infty$.
\hfill \textbf{q.e.d.}	

Since the mapping $\lambda \mapsto \mathrm{P}_\lambda[X=k]$ is continuous, Lemma~\ref{le:diff}
implies that the neutral Bayesian
estimator $\lambda_2^\ast(u)$ is differentiable with respect to $u$ also under Assumption~\ref{as:OneCorr},
with derivative
\begin{equation}\label{eq:corr.increasing}
	\frac{d\,\lambda_2^\ast(u)}{d\,u}\ =\ \mathrm{P}_u[X=k]\,
	\frac{\int_0^u (u-\lambda)\,\mathrm{P}_\lambda[X=k]\,d\,\lambda}%
	{\Big(\int_0^u \mathrm{P}_\lambda[X=k]\,d\,\lambda\Big)^2}\ > \ 0.
\end{equation}
Hence the neutral Bayesian estimator $\lambda_2^\ast(u)$ from Propositions~\ref{pr:correlated} 
is increasing in $u$, as in the independent case.

\begin{table}[t!p]
\caption{One-period, correlated case for different 
asset correlation values. PD estimates under
Assumption~\ref{as:OneCorr} with $k=1$. Naive estimator is $\frac k n$.
Upper confidence bounds and neutral and conservative Bayesian estimators 
according to Proposition~\ref{pr:correlated}.}
\label{tab:2}
\begin{center}
\begin{tabular}{|l|c|c|c|c|c|}
\hline
Estimator & $n=125$ & 250 & 500 & 1000 & 2000 \\ \hline \hline
Naive & 0.8\% & 0.4\% & 0.2\% & 0.1\% & 0.05\% \\ \hline\hline
\multicolumn{6}{|c|}{$\varrho = 0$}\\ \hline\hline
\multicolumn{6}{|c|}{See Table~\ref{tab:1}.} \\ \hline\hline
\multicolumn{6}{|c|}{$\varrho = 0.18$}\\ \hline\hline
50\% upper confidence bound & 2.172\% & 1.213\% & 0.6752\% & 0.3789\% & 0.2101\% \\ \hline
75\% upper confidence bound & 4.6205\% & 2.7141\% & 1.5935\% & 0.9371\% & 0.5494\% \\ \hline
90\% upper confidence bound & 8.3234\% & 5.1456\% & 3.166\% & 1.9408\% & 1.1889\% \\ \hline
Neutral Bayesian on (0, 0.01) & 0.5893\% & 0.5555\% & 0.5146\% & 0.4673\% & 0.4145\% \\ \hline
Neutral Bayesian on (0, 0.1) & 3.747\% & 2.9483\% & 2.2161\% & 1.6063\% & 1.136\% \\ \hline
Neutral Bayesian on (0, 0.25) & 5.1849\% & 3.6091\% & 2.4817\% & 1.701\% & 1.1664\% \\ \hline
Neutral Bayesian on (0,1) & 5.3717\% & 3.6534\% & 2.491\% & 1.7028\% & 1.1669\% \\ \hline
Conservative Bayesian & 5.6706\% & 3.8092\% & 2.5724\% & 1.7455\% & 1.1894\% \\ \hline\hline 
\multicolumn{6}{|c|}{$\varrho = 0.24$}\\ \hline\hline
50\% upper confidence bound & 2.5847\% & 1.4981\% & 0.871\% & 0.5069\% & 0.2939\% \\ \hline
75\% upper confidence bound & 5.7816\% & 3.5573\% & 2.1841\% & 1.3431\% & 0.8216\% \\ \hline
90\% upper confidence bound & 10.7333\% & 6.9794\% & 4.5195\% & 2.9129\% & 1.8711\% \\ \hline
Neutral Bayesian on (0, 0.01) & 0.5909\% & 0.5631\% & 0.5312\% & 0.4955\% & 0.4564\% \\ \hline
Neutral Bayesian on (0, 0.1) & 4.1485\% & 3.5018\% & 2.8692\% & 2.287\% & 1.7805\% \\ \hline
Neutral Bayesian on (0, 0.25) & 6.4935\% & 4.9115\% & 3.6527\% & 2.6923\% & 1.977\% \\ \hline
Neutral Bayesian on (0,1) & 7.1128\% & 5.1411\% & 3.7339\% & 2.7193\% & 1.9855\% \\ \hline
Conservative Bayesian & 7.6721\% & 5.4633\% & 3.9248\% & 2.8324\% & 2.0527\% \\ \hline 
\end{tabular}
\end{center}
\end{table}
Table~\ref{tab:2}, when compared to Table~\ref{tab:1}, shows that the impact of
correlation on the one-period PD estimates is huge. For larger portfolio sizes and
higher confidence levels, the impact of correlation is stronger than for smaller 
portfolios and lower confidence levels. 

While, thanks to the Bayesian estimators, it is possible to get rid of the 
subjectivity inherent by the choice of a confidence level, it is not clear 
how to decide what should be the right level of correlation for the PD estimation.
The values $\varrho = 0.18$ and $\varrho = 0.24$ used for the calculations for
Table~\ref{tab:2} are choices suggested by the Basel~II Accord where the range
of the asset correlation for corporates is defined as $[0.12, 0.24]$. Hence in Table~\ref{tab:2}
we have looked at the mid-range and upper threshold values of the correlation but there
is no convincing rationale of why these values should be more appropriate than others.

The next section explores how to estimate the asset correlation,
while at the same time we extend the range of the estimation samples to time
series of default observations. Clearly, the assumption of having a time series
of default observations for the PD estimation is more realistic than the one-period
models we have studied so far.


\section*{Multi-period observations, correlated defaults}
\label{sec:MultiDep}

According to paragraph 463 of \citet{BaselAccord}, banks applying the IRB approach
    have to use at least five years of historical default data for
    their PD estimations. Ideally, the time series would cover at least one full credit cycle.
Obviously, this requirement calls for a multi-period approach to PD estimation.
    
The portfolio characteristic of low default numbers often can be observed over many years.
Clearly, multiple years of low default numbers should be reflected in the PD estimates.
However, when modelling for multi-period estimation of PDs dependencies over time must be regarded 
    because the portfolio includes the same borrowers over many years and the systemic
    factors causing cross-sectional correlation of default events in different years
    are unlikely to be uncorrelated.
    
In non-technical terms, the framework for the PD estimation methods described in this section
can be explained as follows:    
 \begin{itemize}
\item There is a \textbf{time series} $(n_1, k_1), \ldots, (n_T, k_T)$ of
\begin{itemize}
	\item annual pool sizes $n_1, \ldots, n_T$ (as at the beginning of the year), and
	\item annual observed numbers of defaults $k_1, \ldots, k_T$ (as at the end of the year).
\end{itemize}
\item The pool of borrowers observed for potential default is \textbf{homogeneous} with regard
to the long-run and instantaneous  (point-in-time) PDs:
\begin{itemize}
	\item At a fixed moment in time, all borrowers in the pool have the same instantaneous PD.
	\item All borrowers have the same long-run average PD.
\end{itemize}
\item There is \textbf{dependence} of the borrowers' default behaviour causing 
cross-sectional and over-time default correlation:
\begin{itemize}
	\item At a fixed moment in time, a borrower's instantaneous PD is impacted by an idiosyncratic factor
	and a single systemic factor common
	to all borrowers.
	\item The systemic factors at different moments in time are the more dependent, the
	less the time difference is. 
\end{itemize} 
\end{itemize}
   
The following assumption provides the details of a technical framework for multi-period modelling of portfolio
defaults in the presence of cross-sectional and over time dependencies that has the
afore-mentioned features.

\begin{assumption}\label{as:Multi}
The estimation sample is given by a time series $(n_1, k_1), \ldots, (n_T, k_T)$ of
annual pool sizes $n_1, \ldots, n_T$ and
annual observed numbers of defaults $k_1, \ldots, k_T$ with $0 \le k_1 < n_1$, $\ldots$,
$0 \le k_T < n_T$.\\
All defaults of borrowers have
the same probability of default (PD) parameter $0 < \lambda < 1$. 
Default events at time $t$ are impacted by the \emph{systemic factor} $S_t$ which is assumed to be standard 
normally distributed.\\
The systemic factors $(S_1, \ldots, S_T)$ are jointly normally distributed. 
The correlation of $S_t$ and $S_\tau$ decreases with increasing difference of $t$ and $\tau$
as described in Equation \eqref{eq:tau}:
\begin{subequations}
\begin{equation}\label{eq:tau}
	\mathrm{corr}[S_t, \,S_\tau] \ = \ \vartheta^{|t-\tau|}.
\end{equation}
Default of borrower $A$ occurs at time $t$ if
\begin{equation}\label{eq:trigger}
	\sqrt{\varrho}\,S_t + \sqrt{1-\varrho}\,\xi_{A,\,t} \ \le\ \Phi^{-1}(\lambda).
\end{equation}
\end{subequations}
	Here $\xi_{A,\,t}$ is another standard normal variable, called \emph{idiosyncratic factor}, independent of 
	the idiosyncratic factors relating to the other borrowers and $(S_1, \ldots, S_T)$. \\
The correlation parameters $0 \le \varrho < 1$ and $0 \le \vartheta < 1$ are the same for all 
borrowers and pairs of borrowers respectively.
\end{assumption}
The purpose of the \emph{time-correlation parameter} $\vartheta$ is to capture time-clustering of default observations. 	
By \eqref{eq:tau} the correlation matrix $\Sigma_\vartheta$ of the systemic factors has the following shape:
\begin{equation}\label{eq:Sigma}
	\Sigma_\vartheta \ = \
     \left(%
\begin{array}{ccccc}
  1 & \vartheta & \vartheta^2 & \cdots & \vartheta^{T-1}\\
  \vartheta & 1 & \vartheta & \cdots & \vartheta^{T-2} \\
  \vdots &  & \ddots &  & \vdots \\
  \vartheta^{T-2} & \cdots & \vartheta & 1 & \vartheta \\
  \vartheta^{T-1} & \cdots & \vartheta^2 & \vartheta & 1 
\end{array}%
\right).
\end{equation}
Since the correlation of a pair of systemic factors falls exponentially with 
increasing time difference the dependence structure has a local, short-term character.

As in the previous section, the parameter $\varrho$ is called \emph{asset correlation}. 
It controls the sensitivity of the default events to 
the systemic factors. The larger $\varrho$, the stronger the dependence between different borrowers.

\begin{proposition}\label{pr:multiDist}
\begin{subequations}
Under Assumption~\ref{as:Multi}, denote by $X_t$ the random number of defaults observed in year $t$.
Define the function $G$ by \eqref{eq:G}.\\
Then the distribution of $X_t$ is correlated binomial, as specified by \eqref{eq:corrbinomial}.\\
A borrower's unconditional \emph{(long-run)} probability of default at time $t$ is $\lambda$, i.e.
\begin{equation}\label{eq:uncond}
	\mathrm{P}_\lambda[\text{Borrower $A$ defaults at time $t$}] \ = \ \lambda.
\end{equation}
A borrower's probability of default at time $t$ conditional on a realisation of the 
systemic factors $(S_1, \ldots, S_T)$ \emph{(point-in-time PD)} is given by
\begin{equation}\label{eq:PDcond}
	\mathrm{P}_\lambda[\text{Borrower $A$ defaults at time $t$}\,|\,S_1, \ldots, S_T]\ = \
		G(\lambda,\,\varrho, S_t).
\end{equation}
The probability to observe $k_1$ defaults at time $1$, $\ldots$, $k_T$ defaults at time $T$,
conditional on a realisation of the 
systemic factors $(S_1, \ldots, S_T)$ is given by
\begin{equation}\label{eq:condLik}
	\mathrm{P}_\lambda[X_1 = k_1, \ldots, X_T = k_T\,|\,S_1, \ldots, S_T] \,= \, 
	\prod_{t=1}^T \left(\begin{smallmatrix} n_t\\ k_t \end{smallmatrix}\right) 
	G(\lambda,\,\varrho, S_t)^{k_t}\bigl(1-G(\lambda,\,\varrho, S_t)\bigr)^{n_t-k_t}.
\end{equation}
The unconditional probability to observe $k_1$ defaults at time $1$, $\ldots$, $k_T$ defaults at time $T$
is given by
\begin{multline} \label{eq:uncondLik}
	\mathrm{P}_\lambda[X_1 = k_1, \ldots, X_T = k_T] \ = \
	\idotsint \varphi_{\Sigma_\vartheta}(s_1, \ldots, s_T) \\ \prod_{t=1}^T \left(\begin{smallmatrix} n_t\\ k_t \end{smallmatrix}\right) 
	G(\lambda,\,\varrho, S_t)^{k_t}\bigl(1-G(\lambda,\,\varrho, S_t)\bigr)^{n_t-k_t}\,
	d(s_1, \ldots, s_T),
\end{multline}
where $\varphi_{\Sigma_\vartheta}$ denotes the multi-variate normal density 
 with mean 0 and covariance 
matrix $\Sigma_\vartheta$ as defined by \eqref{eq:Sigma} (see, e.g., Section 3.1.3 of
\citet{McNeil05} for the formal definition).
\end{subequations}
\end{proposition}
\textbf{Proof.} For a fixed time $t$ Assumption~\ref{as:Multi} implies Assumption~\ref{as:OneCorr}. 
By Proposition~\ref{pr:corrbinomial}, this implies that $X_t$ is correlated binomial and 
\eqref{eq:uncond}. By independence of $\xi_A$ and $(S_1, \ldots, S_T)$ and the fact that
$\xi_{A,t}$ is standard normal, \eqref{eq:PDcond} follows from \eqref{eq:trigger}. 
Equation~\eqref{eq:condLik} follows from the observation that the default events as
specified by \eqref{eq:trigger} are independent conditional on realisations of the
systemic factors $(S_1, \ldots, S_T)$. Equation~\eqref{eq:uncondLik} is then
an immediate consequence of the definition of conditional probability. \hfill \textbf{q.e.d.}

\begin{remark}
By \eqref{eq:uncondLik}, for $\lambda > 0$, there is a positive -- if very small --
probability of observing $n_1 + n_2 + \ldots + n_T$ defaults during the observation
period of $T$ years. However, in realistic portfolios this event would be impossible and
hence have probability zero.

This observation implies that Assumption~\ref{as:Multi} is not fully realistic. It is possible
to make Assumption~\ref{as:Multi} more realistic by providing exact information about the years
each borrower spent in the portfolio and about the reasons why borrowers disappeared from the
portfolio (default or regular termination of the transactions with the borrower). 

The original method for multi-period low default estimation suggested by \citet{Pluto&Tasche}
is based on such a cohort approach. \citet{Pluto&Tasche} actually considered only the case
where a cohort of borrowers being in the portfolio at time $1$ was observed over time, without
the possibility to leave the portfolio regularly. In addition, \citeauthor{Pluto&Tasche} assumed
that no new borrowers entered the portfolio. This latter assumption can be removed, but at high
computational cost.

In this paper, we focus on the simpler (but slightly unrealistic) approach developed on the basis
of Assumption~\ref{as:Multi} and Proposition~\ref{pr:multiDist}. This approach was called 
\emph{multiple binomial} in \citet{Pluto&Tasche2011} and its numerical results were compared to
results calculated by means of the \emph{cohort approach} from \citet{Pluto&Tasche}. \citeauthor{Pluto&Tasche}
found that the differences of the results by the two approaches were negligible. Thus, the
multiple binomial approach based on Assumption~\ref{as:Multi} can be considered a reasonable
approximation to the more realistic but also more involved cohort approach.
\end{remark}

In principle, both \eqref{eq:condLik} and \eqref{eq:uncondLik} can serve as the
basis for maximum likelihood estimation of the model parameters $\lambda$ (PD),
$\varrho$ (asset correlation), and $\vartheta$ (time correlation). Using \eqref{eq:condLik}
for maximum likelihood estimation requires the identification of the systemic factors
with real, observable economic factors that explain all the systemic risk of the default
events. While for corporate portfolios there are promising candidates for the identification
of the systemic factors (see \citet{Aguais&al2006} for an example),
it is not clear whether it is indeed possible to explain all the systemic risk of
the portfolios by the time evolution of just one observable factor. Moreover, there
are low default portfolios like banks or public sector entities for which there are no 
obvious observable economic factors that are likely to explain most of the systemic
risk of the portfolios.

In the following, it is assumed that the systemic factors $(S_1, \ldots, S_T)$ are latent
(not observable) and that, hence, maximum likelihood estimation of the model parameters
$\lambda$, $\varrho$, and $\vartheta$ must be based on Equation~\eqref{eq:uncondLik}. The
right-hand side of \eqref{eq:uncondLik} is then proportionate to 
the marginal likelihood function that must be 
maximised as a function of the model parameters. In technical terms, the related
procedure for finding the \emph{maximum likelihood estimates} 
$\hat{\lambda}$, $\hat{\varrho}$, and $\hat{\vartheta}$ can be described as
\begin{multline} \label{eq:likelihood}
(\hat{\lambda}, \hat{\varrho}, \hat{\vartheta}) \ = \
\arg\max_{(\lambda, \varrho, \vartheta)}	\idotsint 
\varphi_{\Sigma_\vartheta}(s_1, \ldots, s_T) \\ 
\prod_{t=1}^T 
	G(\lambda, \varrho, s_t)^{k_t}\bigl(1-G(\lambda, \varrho, s_t)\bigr)^{n_t-k_t}
	d(s_1, \ldots, s_T).
\end{multline}
Solving the optimization problem \eqref{eq:likelihood} is demanding as it involves multi-dimensional
integration and the determination of an absolute maximum with respect to three variables.
For Example~\ref{ex:fictitious} and Example~\ref{ex:Moodys} below, 
the multiple integrals were calculated by means of Monte-Carlo
simulation while the procedure \emph{nlminb} from the software package R \citep{RSoftware}
was applied to the optimization problem.
Note that the maximum likelihood estimates of $\lambda$, $\varrho$, and $\vartheta$ are 
different from 0 only if $k_1 + \ldots + k_T > 0$ (i.e.\ 
only if at least one default was observed). 
	
Maximum likelihood estimates are best estimates in some sense but are not necessarily conservative.
In particular, if there are no default observations the maximum likelihood estimate
of the long-run PD is zero -- which is unsatisfactory from the perspective of prudent 
risk management. That is why it makes sense to extend the upper confidence bound and Bayesian
approaches from the previous two sections to the multi-period setting
as described by Assumption~\ref{as:Multi}.
Bayesian estimates in the context of Assumption~\ref{as:Multi} are straight-forward while
the determination of upper confidence bounds requires another approximation since convolutions of
binomial distributions are not binomially but at best approximately Poisson distributed.

\begin{proposition}\label{pr:multiBayesian}
Under Assumption~\ref{as:Multi}, denote by $X_t$ the random number of defaults observed in year $t$.
Let $\mathrm{P}_\lambda[X_1 = k_1, \ldots, X_T = k_T]$ be given by \eqref{eq:uncondLik} and
let $X = X_1 + \ldots + X_T$ denote the total number of defaults observed in the time period from
$t=1$ to $t=T$. Define the function $G$ by \eqref{eq:G} and let $k = k_1 + \ldots + k_T$.\\
Then we have the following estimators for
the PD parameter $\lambda$:
\begin{itemize}
	\item[(i)] 
For any fixed confidence level $0 < \gamma < 1$, the upper confidence bound
$\lambda_0^\ast(\gamma)$ for the PD $\lambda$ at level $\gamma$ can be approximately calculated 
by solving the following equation for $\lambda$:
\begin{subequations}
	\begin{align} 
	1 - \gamma & \ =\ \mathrm{P}_\lambda[X \le k] \notag\\ 
	& \ \approx \ \idotsint \varphi_{\Sigma_\vartheta}(s_1, \ldots, s_T) 
	\exp(- I_{\lambda,\,\varrho}(s_1, \ldots, s_T)) \notag\\
	& \qquad\qquad  \sum_{j=0}^k 
	\frac{I_{\lambda,\,\varrho}(s_1, \ldots, s_T)^j}{j!}\,	d(s_1, \ldots, s_T), \label{eq:poisson}\\
	I_{\lambda,\,\varrho}(s_1, \ldots, s_T) & \ =\ \sum_{t=1}^T n_t\,G(\lambda,\,\varrho, s_t).\notag
\end{align}			
\item[(ii)] If the Bayesian prior distribution of the PD $\lambda$ is given by \eqref{eq:apriori}
then the mean $\lambda_1^\ast$ of the posterior distribution is given by
\begin{equation}\label{eq:posterior.conservative}
	\lambda_1^\ast \ =\ \frac{\int_0^1 \frac{\lambda\,\mathrm{P}_\lambda[X_1 = k_1, \ldots, X_T = k_T]}
	{1-\lambda}\, d\lambda}%
			{\int_0^1 \frac{\mathrm{P}_\lambda[X_1 = k_1, \ldots, X_T = k_T]}{1-\lambda}\, d\lambda}.	
\end{equation}
In particular, the integrals in the numerator and the denominator of the
right-hand side of \eqref{eq:posterior.conservative} are finite.
$\lambda_1^\ast$ is called the \emph{conservative Bayesian estimator} of the PD $\lambda$. 
\item[(iii)] If the Bayesian prior distribution of the PD $\lambda$ is 
uniform on $(0,u)$ for some $0 < u \le 1$ then the mean $\lambda_2^\ast(u)$ 
of the posterior distribution is given by
\begin{equation}\label{eq:posterior.neutral}
	\lambda_2^\ast(u) \ =\ \frac{\int_0^u \lambda\,\mathrm{P}_\lambda[X_1 = k_1, \ldots, X_T = k_T]\, d\lambda}%
			{\int_0^u \mathrm{P}_\lambda[X_1 = k_1, \ldots, X_T = k_T]\, d\lambda}.
\end{equation}
$\lambda_2^\ast(u)$ is called the \emph{$(0,u)$-constrained neutral Bayesian estimator} of the PD $\lambda$. 
For $u = 1$, we obtain the (unconstrained) \emph{neutral Bayesian estimator} $\lambda_2^\ast(1)$.
\end{subequations}
\end{itemize}
\end{proposition}
\textbf{Proof.} As the $X_t$ are independent and binomially distributed conditional on
realizations of the systemic factors $(S_1, \ldots, S_T)$, they are approximately Poisson distributed
conditional on $(S_1, \ldots, S_T)$, with an intensities $n_t\,G(\lambda,\,\varrho, s_t)$, $t=1, \ldots, T$.
Approximation \eqref{eq:poisson} follows because the sum of independent Poisson distributed variables 
is again Poisson distributed, with intensity
equal to the sum of the intensities of the variables.
Formulae \eqref{eq:posterior.conservative} and \eqref{eq:posterior.neutral} for
the Bayesian estimators are straightforward. The finiteness of the integrals on the right-hand side of
\eqref{eq:posterior.conservative} can be shown as in the proof of Proposition~\ref{pr:correlated}.
\hfill \textbf{q.e.d.}

Observe that $\mathrm{P}_\lambda[X_1 = k_1, \ldots, X_T = k_T]$ as given by \eqref{eq:uncondLik} is
continuous in $\lambda$. By Lemma~\ref{le:diff} this implies that $u \mapsto \lambda_2^\ast(u)$ is
increasing with $u$ also under Assumption~\ref{as:Multi}, again as is to be intuitively expected.

We are going to illustrate the multi-period estimators of the correlation parameters and the PD $\lambda$ 
that have been presented in \eqref{eq:likelihood}
and in Proposition~\ref{pr:multiBayesian} by two numerical examples. The first of the examples is for
comparison with the results in Tables~\ref{tab:1} and \ref{tab:2} and, therefore, is fictitious.
The second example is based on real default data as reported by \citet{Moodys2011}. Before we present
the examples, it is worthwhile to provide some comments on the numerical calculations needed
for the evaluation of the estimators.

The main difficulty in the numerical calculations for the multi-period setting is the evaluation
of the unconditional probability \eqref{eq:uncondLik} as it requires multi-dimensional integration.
For the purpose of this paper, we approximate the multi-variate integral by means of Monte-Carlo simulation,
i.e.\ we generate a sample $(s_1^{(1)}, \ldots, s_T^{(1)})$, $\ldots$, $(s_1^{(n)}, \ldots, s_T^{(n)})$ of
independent realisations of the jointly normally distributed systemic factors $(S_1, \ldots, S_T)$
from Assumption~\ref{as:Multi} and compute
\begin{equation} \label{eq:multi.approx}
	\mathrm{P}_\lambda[X_1 = k_1, \ldots, X_T = k_T] \ \approx \
	1/n \sum_{i=1}^n  \prod_{t=1}^T \left(\begin{smallmatrix} n_t\\ k_t \end{smallmatrix}\right) 
	G(\lambda,\,\varrho, s_t^{(i)})^{k_t}\bigl(1-G(\lambda,\,\varrho, s_t^{(i)})\bigr)^{n_t-k_t}.
\end{equation}
The right-hand side of \eqref{eq:poisson} is similarly approximated. The estimators \eqref{eq:posterior.conservative}
and \eqref{eq:posterior.neutral}, however, require an additional integration with respect to a uniformly distributed
variable. With a view of preserving the monotonicity property of $u \mapsto \lambda_2^\ast(u)$ and 
efficient calculation of $\lambda_2^\ast(u)$ for different $u$ we approximate the estimators $\lambda_1^\ast$
and $\lambda_2^\ast(u)$ in the following specific way that might not be most efficient. 

For fixed $0 < u \le 1$ choose a positive integer $m$ and let
\begin{subequations}
\begin{equation}
	u_i \ =\ \frac i m \,u, \quad i = 0, 1, \ldots, m.
\end{equation}
Generate a sample $(s_1^{(1)}, \ldots, s_T^{(1)})$, $\ldots$, $(s_1^{(n)}, \ldots, s_T^{(n)})$ of
independent realisations of the jointly normally distributed systemic factors $(S_1, \ldots, S_T)$
from Assumption~\ref{as:Multi}, with $n$ being an integer possibly different to $m$. Based on 
$(u_0, \ldots, u_m)$ and $(s_1^{(1)}, \ldots, s_T^{(1)})$, $\ldots$, $(s_1^{(n)}, \ldots, s_T^{(n)})$
we then use the below estimators of $\lambda_1^\ast$ and $\lambda_2^\ast(u)$:
\begin{align}\label{eq:lambda1}
	\lambda_1^\ast &\ \approx\ \frac{\sum_{i=0}^{m-1} u_i\,(1-u_i)^{-1}\sum_{j=1}^n  \prod_{t=1}^T 
	G(u_i,\,\varrho, s_t^{(j)})^{k_t}\bigl(1-G(u_i,\,\varrho, s_t^{(j)})\bigr)^{n_t-k_t}}%
	{\sum_{i=0}^{m-1} (1-u_i)^{-1}\sum_{j=1}^n  \prod_{t=1}^T 
	G(u_i,\,\varrho, s_t^{(j)})^{k_t}\bigl(1-G(u_i,\,\varrho, s_t^{(j)})\bigr)^{n_t-k_t}},\\[1ex]
	\lambda_2^\ast(u) &\ \approx\ \frac{\sum_{i=0}^m u_i \sum_{j=1}^n  \prod_{t=1}^T 
	G(u_i,\,\varrho, s_t^{(j)})^{k_t}\bigl(1-G(u_i,\,\varrho, s_t^{(j)})\bigr)^{n_t-k_t}}%
	{\sum_{i=0}^m \sum_{j=1}^n  \prod_{t=1}^T 
	G(u_i,\,\varrho, s_t^{(j)})^{k_t}\bigl(1-G(u_i,\,\varrho, s_t^{(j)})\bigr)^{n_t-k_t}}.\label{eq:lambda2}
\end{align}
\end{subequations}
The right-hand side of \eqref{eq:lambda1} has been stated deliberately for general $u \le 1$ although
in theory according to \eqref{eq:posterior.conservative} only $u = 1$ is needed. The reason for
this generalization is that the values of the functions integrated 
in \eqref{eq:posterior.conservative} and \eqref{eq:posterior.neutral}
are very close to zero for $\lambda$ much greater than $\frac{\sum_{t=1}^T k_t}{\sum_{t=1}^T n_t}$ 
and, therefore, can be ignored for the purpose of evaluating the integrals.

\begin{table}[t!p]
\caption{Fictitious default data for Example~\ref{ex:fictitious}.}
\label{tab:3}
\begin{center}
\begin{tabular}[t]{|c|c|c|}\hline
Year & Pool size & Defaults\\ \hline\hline
2003 &	125 &	0 \\ \hline
2004 &	125 &	0 \\ \hline
2005 &	125 &	0 \\ \hline
2006 &	125 &	0 \\ \hline
2007 & 	125 &	0 \\ \hline
2008 &	125 &	0 \\ \hline
2009 &	125 &	0 \\ \hline
2010 & 	125 &	1 \\ \hline\hline
All & 	1000 &	1 \\ \hline
\end{tabular}
\end{center}
\end{table}

\begin{example}[Fictitious data]\label{ex:fictitious}
We apply the estimators \eqref{eq:likelihood}, \eqref{eq:poisson}, \eqref{eq:posterior.neutral},
and \eqref{eq:posterior.conservative} to the fictitious default data time series presented in 
Table~\ref{tab:3}. The output generated by the calculation with an R-script is listed in
Appendix~\ref{sec:appA}.
\end{example}
\begin{example}[Real data]\label{ex:Moodys}
We apply the estimators \eqref{eq:likelihood}, \eqref{eq:poisson}, \eqref{eq:posterior.neutral},
and \eqref{eq:posterior.conservative} to the default data time series presented in 
Table~\ref{tab:4} in order to determine a long-run PD estimate for entities rated
as investment grade (grades Aaa, Aa, A, and Baa) by the rating agency Moody's. 
The output generated by the calculation with an R-script is listed in
Appendix~\ref{sec:appB}.
\end{example}

Comments on the computation characteristics and results shown in Appendices~\ref{sec:appA}
and \ref{sec:appB}:
\begin{itemize}
	\item The calculation output documented in both appendices starts with some characteristics
of the Monte Carlo simulations used in the course of the calculations. The computations for 
the two maximum likelihood (ML)
estimators (for the three parameters $\lambda$, $\varrho$, and $\vartheta$ together and for $\lambda$
alone, with pre-defined values of $\varrho$ and $\vartheta$) are based on 16 runs of 10,000 
iterations, effectively producing estimates each based on 160,000 iterations. Similarly,
the computations for the upper confidence bounds are each based
on 16 runs of 10,000 iterations. 
	\item Sixteen Monte Carlo runs were also used for the Bayesian estimators.
However, as the Bayesian estimation according to Proposition~\ref{pr:multiBayesian} 
requires inner integration for the unconditional probabilities and
outer integration with respect to $\lambda$ the documented calculation output lists both the number
of simulation iterations ($n$ in \eqref{eq:posterior.conservative} and
\eqref{eq:posterior.neutral}) for the inner integral and the number of steps 
($m$ in \eqref{eq:posterior.conservative} and
\eqref{eq:posterior.neutral}) in the outer integral.
	\item The split into
16 runs was implemented in order to deliver rough estimates of the estimation uncertainty
inherent in the Monte Carlo simulation. The standard deviations shown in Appendices~\ref{sec:appA}
and \ref{sec:appB} below the different estimates are effectively the standard deviations
of the means of the 16 runs each with 10,000 iterations. Hence, the standard deviation
of a single run of 10,000 iterations can be determined by multiplying the tabulated 
standard deviations with $4 = \sqrt{16}$.
	\item Below the Monte Carlo characteristics, summary metrics of the default data from Tables~\ref{tab:3}
	and \ref{tab:4} respectively are shown. The naive PD estimates are calculated as the number
	of observed defaults divided by number of obligor-years.
	\item The maximum likelihood estimates listed in the appendices were determined by 
	solving the optimisation problem~\eqref{eq:likelihood} (case of estimated correlations)
	and the related optimisation problem for the PD $\lambda$ only (case of pre-defined
	correlations). The calculations for the upper confidence bounds and the Bayesian estimators
	were based on the formulae presented in Proposition~\ref{pr:multiBayesian}. In addition,
	Monte Carlo approximations according to \eqref{eq:multi.approx}, \eqref{eq:lambda1} and
	\eqref{eq:lambda2} were used.
	\item In both cases (estimated correlations and pre-defined correlations respectively)
	 the (unconstrained) neutral and conservative Bayesian estimates were approximated
	by the $(0, 0.1)$-constrained estimates (i.e.\ $u = 0.1$ in \eqref{eq:lambda1} and
	\eqref{eq:lambda2}). Test calculations not
	documented in this paper showed that there is practically no difference between these
	constrained estimates and the unconstrained estimates (with $u = 1$)
	as long as the naive estimates are of a magnitude of not more than a few basis 
	points.	
	\item The
	constrained neutral Bayesian estimates were calculated with the constraint $u$ given 
	by the corresponding 99\%-upper confidence bounds of the long-run PD parameter $\lambda$.
\end{itemize}

Some observations on the estimation results for Examples~\ref{ex:fictitious} and \ref{ex:Moodys}
as presented in Appendices~\ref{sec:appA} and \ref{sec:appB}:
\begin{itemize}	
	\item The multi-period case is situated between the independent and correlated one-period cases,
in the sense of exhibiting heavier tails of the default number distribution than
the independent one-period case and lighter tails of the default number distribution than
the correlated one-period case. This follows from a comparison of the upper confidence bound
results in Appendix~\ref{sec:appA} to the $n=1000$ columns in Tables~\ref{tab:1} and
\ref{tab:2}. In general, the heavier the tail of the default number
distribution is as a consequence of default correlation,
the stronger is the growth of the upper confidence bounds with increasing
confidence level. 
	\item Example~\ref{ex:fictitious}, case ``estimated correlations'', and Example~\ref{ex:Moodys}, both
	cases of estimated and pre-defined correlations, are qualitatively closer to the one-period independent case while
	Example~\ref{ex:fictitious} ``pre-defined correlations'' is qualitatively closer to the 
	one-period correlated case.
	\item In Example~\ref{ex:fictitious}, case ``estimated correlations'', the estimated asset correlation 
	is zero. Therefore, the case ``estimated correlations'' of Example~\ref{ex:fictitious}  is
	indeed equivalent to the independent one-period case (cf.\ results for upper bounds
	and unconstrained neutral Bayesian estimate in Appendix~\ref{sec:appA} and the column
	for portfolio size $n = 1000$ in Table~\ref{tab:1}).
	\item Due to the relatively long time span of 21 years covered by the default data time series
	in Example~\ref{ex:Moodys}, the average of the elements in the time correlation matrix
	\eqref{eq:Sigma} is quite small. As the portfolio sizes in any fixed year are small
	compared to the number of all observations in the time series, therefore, the tail
	of the default number distribution is relatively light, as in the independent one-period case. 
	This is, in particular, indicated in the results shown in Appendix~\ref{sec:appB} by the fact that
	the Bayesian estimates for all three cases (unconstrained neutral, constrained neutral, 
	conservative) are practically identical.
	\item In the case ``pre-defined correlations'' of Example~\ref{ex:fictitious} the combination
	of a relatively short time series with significant asset correlation and time correlation 
	triggers a relatively heavy-tailed default number distribution. One consequence of this are the
	significant differences between the three different Bayesian estimates presented in 
	Appendix~\ref{sec:appA}. This behaviour is more similar to the behaviour of the Bayesian
	estimators in the one-period correlated case (see Table~\ref{tab:2}) than to the behaviour
	of these estimators in the one-period independent case (see Table~\ref{tab:1}).
	\item The conservative Bayesian estimates for small portfolio sizes and higher default
	correlation are significantly greater than the neutral Bayesian estimates while in the case of
	independent defaults there is hardly any difference between the conservative and the neutral
	estimates.	
	\item Undocumented observation: Due to the portfolio size
	and the length of the time series, Example~\ref{sec:appB} is close to the limits of what can still be dealt with
	by means of the numerical procedures described in this paper.
	See \citet{Wilde&Jackson} for an alternative, less computationally intensive approach to
	the calculations.
\end{itemize}

The observations on Example~\ref{ex:fictitious} and \ref{ex:Moodys} suggest that
the neutral Bayesian estimator (applied as $(0, 0.1)$-constrained estimator)
 gives appropriately conservative
estimates of the long-run PD parameter (not only in the very 
	low default case). This estimator generates estimates between the 50\% and 75\% upper confidence bounds in
	the less correlated cases (short time series with low asset correlation or longer time
	series) and estimates between the 75\% and 90\% upper confidence bounds in the more
	correlated cases. This way, the neutral Bayesian estimator is more sensitive
	to the presence of correlation in the data than the upper confidence bound estimators.
	The conservative Bayesian estimator has a similar property of being sensitive to
	correlation but appears not to differ significantly from
	the neutral estimator even in the presence of correlation. 

\begin{table}[t!p]
\caption{Default data for Example~\ref{ex:Moodys}.
The table lists the numbers of entities rated as investment grade 
(grades Aaa, Aa, A, and Baa) by Moody's at the beginning of the year and the numbers
of defaults among these entities observed by year end. 
Source: Exhibits 17 and 42 of \citet{Moodys2011}.}
\label{tab:4}
\begin{center}
\begin{tabular}[t]{|c|c|c|}\hline
Year & Pool size & Defaults\\ \hline\hline
1990 & 1492 & 0 \\ \hline
1991 & 1543 & 1 \\ \hline
1992 & 1624 & 0 \\ \hline
1993 & 1731 & 0 \\ \hline
1994 & 1888 & 0 \\ \hline
1995 & 2012 & 0 \\ \hline
1996 & 2209 & 0 \\ \hline
1997 & 2412 & 0 \\ \hline
1998 & 2593 & 1 \\ \hline
1999 & 2742 & 1 \\ \hline
2000 & 2908 & 4 \\ \hline
2001 & 2994 & 4 \\ \hline
2002 & 3128 & 14 \\ \hline
2003 & 3015 & 0 \\ \hline
2004 & 2977 & 0 \\ \hline
2005 & 3025 & 2 \\ \hline
2006 & 3082 & 0 \\ \hline
2007 & 3108 & 0 \\ \hline
2008 & 3133 & 14 \\ \hline
2009 & 3048 & 11 \\ \hline
2010 & 2966 & 2 \\ \hline\hline
All & 53630 & 54 \\ \hline
\end{tabular}
\end{center}
\end{table}


\section*{Conclusions}
\label{sec:Conclusions}

In this paper, we have revisited the approach for the estimation of PDs for low
default portfolios as suggested in \citet{Pluto&Tasche}. For the one-period case
with independent default events, we have shown that the upper confidence bounds
for the PDs can be calculated as quantiles of the Bayesian posterior distribution
for a simple prior that is more conservative than the uninformed neutral prior. This observation
suggests that Bayesian estimators computed as means of the posterior distributions
can serve as an alternative to the upper confidence bounds approach.
Such an alternative is welcome because it makes the necessarily subjective choice
of a confidence level redundant.

We have explored generalisations of the conservative Bayesian estimator of the one-period independent
case for the one- and multi-period correlated cases and compared their estimates 
to estimates by means of upper confidence bounds and of Bayesian estimators with 
constrained and unconstrained neutral priors. We have found that a constrained neutral
Bayesian estimator delivers plausible estimates and is sensitive to the presence of
correlation by being situated between the 50\% and 75\% upper bounds for low correlation regimes
and between the 75\% and 90\% upper bounds for higher correlation regimes. Constrained
neutral Bayesian estimators are computationally more efficient than the unconstrained neutral
Bayesian estimators but good approximations
when the constraints are carefully chosen. In particular, the $(0, 0.1)$-constrained
neutral Bayesian estimator appears to be an appropriate tool for conservative long-run PD estimation,
avoiding the issue of which confidence level to choose for the estimation.



\pagebreak


\appendix

\section{Appendix: Output of calculation for Example~\ref{ex:fictitious}}
\label{sec:appA}

{\small\begin{alltt}
Sun Apr 01 17:01:21 2012 
Multiperiod low default estimation
Fictitious Default Data\vspace{1ex} 
Random seed: 36 
Number of ML simulation iterations: 10000 
Number of ML simulation runs: 16 
Number of confidence bounds simulation iterations: 10000 
Number of confidence bounds simulation runs: 16 
Number of inner Bayesian simulation iterations: 1000 
Number of outer Bayesian steps: 2500 
Number of Bayesian simulation runs: 16 
Length of time period: 8 
Total number of obligor-years: 1000 
Total observed number of defaults: 1 
Naive PD estimate (bps): 10\vspace{1ex} 
Estimates with estimated correlations:
ML estimate for PD (bps):  10.0 
Standard deviation (bps):   0.0 
ML estimate for rho (%):   0.0 
Standard deviation (%):   0.0 
ML estimate for theta (%):  12.4 
Standard deviation (%):   4.7\vspace{1ex} 
Conf. level (%)  & 50.00 & 75.00 & 90.00 & 95.00 & 99.00 & 99.90 
Upper bound (bps) & 16.8 & 26.9 & 38.8 & 47.3 & 66.2 & 92.6 
Std. dev. (bps)   &  0.0 &  0.0 &  0.0 &  0.0 &  0.0 &  0.0\vspace{1ex} 
Bayesian neutral estimate for PD (bps):  20.0 
Standard deviation (bps):   0.0 
Bayesian constrained estimate for PD (bps):  19.4 
Standard deviation (bps):   0.0 
Bayesian conservative estimate for PD (bps):  20.0 
Standard deviation (bps):   0.0\vspace{1ex} 
Estimates with pre-defined correlations:
Asset correlation (%): 18.0 
Time correlation deployed (%): 60.0 
ML estimate for PD (bps) only:  14.1 
Standard deviation (bps):   0.1\vspace{1ex} 
Conf. level (%)  & 50.00 & 75.00 & 90.00 & 95.00 & 99.00 & 99.90 
Upper bound (bps) & 23.5 & 48.3 & 86.4 & 119.4 & 209.4 & 368.9 
Std. dev. (bps)   &  0.3 &  0.5 &  0.9 &  1.1 &  2.6 &  7.7\vspace{1ex} 
Bayesian neutral estimate for PD (bps):  58.7 
Standard deviation (bps):   1.3 
Bayesian constrained estimate for PD (bps):  53.4 
Standard deviation (bps):   0.5 
Bayesian conservative estimate for PD (bps):  61.6 
Standard deviation (bps):   1.1  
\end{alltt}}

\pagebreak

\section{Appendix: Output of calculation for Example~\ref{ex:Moodys}}
\label{sec:appB}

{\small\begin{alltt}
Sun Apr 01 18:38:09 2012 
Multiperiod low default estimation
Moody's Investment Grade\vspace{1ex}
Random seed: 36 
Number of ML simulation iterations: 10000 
Number of ML simulation runs: 16 
Number of confidence bounds simulation iterations: 10000 
Number of confidence bounds simulation runs: 16 
Number of inner Bayesian simulation iterations: 1000 
Number of outer Bayesian steps: 2500 
Number of Bayesian simulation runs: 16 
Length of time period: 21 
Total number of obligor-years: 53630 
Total observed number of defaults: 54 
Naive PD estimate (bps): 10.1\vspace{1ex} 
Estimates with estimated correlations:
ML estimate for PD (bps):  17.6 
Standard deviation (bps):   2.5 
ML estimate for rho (%):  24.3 
Standard deviation (%):   1.2 
ML estimate for theta (%):  58.0 
Standard deviation (%):   4.6\vspace{1ex} 
Conf. level (%)  & 50.00 & 75.00 & 90.00 & 95.00 & 99.00 & 99.90 
Upper bound (bps) & 14.3 & 23.6 & 35.7 & 45.2 & 69.5 & 109.5 
Std. dev. (bps)   &  0.2 &  0.3 &  0.3 &  0.5 &  1.3 &  6.2\vspace{1ex} 
Bayesian neutral estimate for PD (bps):  16.6 
Standard deviation (bps):   2.2 
Bayesian constrained estimate for PD (bps):  16.5 
Standard deviation (bps):   2.2 
Bayesian conservative estimate for PD (bps):  16.6 
Standard deviation (bps):   2.2\vspace{1ex} 
Estimates with pre-defined correlations:
Asset correlation (%): 18.0 
Time correlation deployed (%): 60.0 
ML estimate for PD (bps) only:  11.5 
Standard deviation (bps):   1.4\vspace{1ex} 
Conf. level (%)  & 50.00 & 75.00 & 90.00 & 95.00 & 99.00 & 99.90 
Upper bound (bps) & 12.8 & 20.0 & 29.1 & 36.2 & 52.9 & 79.7 
Std. dev. (bps)   &  0.1 &  0.2 &  0.2 &  0.4 &  1.0 &  3.7\vspace{1ex} 
Bayesian neutral estimate for PD (bps):  15.6 
Standard deviation (bps):   2.3 
Bayesian constrained estimate for PD (bps):  15.6 
Standard deviation (bps):   2.3 
Bayesian conservative estimate for PD (bps):  15.6 
Standard deviation (bps):   2.3 
\end{alltt}}

\end{document}